%% file: 00paper.tex
\documentclass[runningheads]{llncs}
\usepackage{graphicx}
\usepackage{hyperref}
\usepackage{booktabs}
\usepackage{adjustbox}
\usepackage[dvipsnames]{xcolor}
\usepackage[inline]{enumitem}
\usepackage[square,comma,numbers,sort&compress,sectionbib]{natbib}
\usepackage{fancyhdr}

\fancypagestyle{specialfooter}{%
  \fancyhf{}
  
  \fancyfoot[C]{\vspace{1cm} \hspace*{-.2\textwidth}\parbox{1.4\textwidth}{\small This is the author's version of the work. It is posted here for your personal use. The definitive version is published in: \\ \emph{Proceedings of the 46th European Conference on Information Retrieval} (ECIR '24), March 24--28, 2024, Glasgow, Scotland}}
}

\begin{document}
\mainmatter
\title{Estimating the Usefulness of Clarifying Questions and Answers for Conversational Search}
\titlerunning{Usefulness of Clarifying Questions and Answers}

\author{Ivan Sekuli\'c\inst{1}\orcidID{0009-0001-0036-0490} \and
\mbox{Weronika Łajewska}\inst{2}\orcidID{0000-0003-2765-2394} \and
\mbox{Krisztian Balog}\inst{2}\orcidID{0000-0003-2762-721X} \and
\mbox{Fabio Crestani}\inst{1}\orcidID{0000-0001-8672-0700}}
\authorrunning{I. Sekuli\'c et al.}
\institute{Università della Svizzera italiana, Lugano, Switzerland \\
\email{\{ivan.sekulic,fabio.crestani\}@usi.ch},\\
\and University of Stavanger, Stavanger, Norway \\
\email{\{weronika.lajewska,krisztian.balog\}}@uis.ch}

\maketitle

\begin{abstract}
While the body of research directed towards constructing and generating clarifying questions in mixed-initiative conversational search systems is vast, research aimed at processing and comprehending users' answers to such questions is scarce.
To this end, we present a simple yet effective method for processing answers to clarifying questions, moving away from previous work that simply appends answers to the original query and thus potentially degrades retrieval performance.
Specifically, we propose a classifier for assessing usefulness of the prompted clarifying question and an answer given by the user.
Useful questions or answers are further appended to the conversation history and passed to a transformer-based query rewriting module.
Results demonstrate significant improvements over strong non-mixed-initiative baselines.
Furthermore, the proposed approach mitigates the performance drops when non useful questions and answers are utilized.

\end{abstract}

\thispagestyle{specialfooter}

\input{ecir2023-castmi-01}

\input{ecir2023-castmi-02}
\input{ecir2023-castmi-03}
\input{ecir2023-castmi-05}
\input{ecir2023-castmi-06}
\input{ecir2023-castmi-07}

\renewcommand*{\bibfont}{\scriptsize}
\bibliographystyle{splncs04}
\bibliography{ecir2023-castmi.bib}
\end{document}

%% file: ecir2023-castmi-01.tex
\section{Introduction}

The goal of a conversational search (CS) system is to satisfy the user's information need.
To this end, several aspects of CS systems have enjoyed significant advancements, including conversational passage retrieval~\cite{Yu:2021:SIGIR}, query rewriting~\cite{Vakulenko:2021:ECIR}, and intent detection~\cite{Qu:2019:CHIIR}.
In recent years, the mixed-initiative paradigm of conversational search has attracted significant research attention~\cite{Aliannejadi:2019:SIGIR,Rao:2018:ACL,Rosset:2020:WWW}.
Under the mixed-initiative paradigm, the CS system can at any point in a conversation offer suggestions or ask clarifying questions.

Asking clarifying questions has been identified as an invaluable component of modern-day CS systems~\cite{Radlinski:2017:CHIIR}.
Such questions are directed at users and aim to elucidate their underlying information needs.
While there is a growing body of research on constructing and generating clarifying questions~\cite{Aliannejadi:2019:SIGIR,Sekulic:2021:ICTIR,Zamani:2020:WWW}, work aimed at processing and comprehending users' answers to such questions is scarce.
Nonetheless, recent research suggests their usefulness by demonstrating improvements in passage retrieval performance after asking a clarifying question and receiving an answer~\cite{Aliannejadi:2021:EMNLP}.

To bridge the aforementioned research gap, we make a first step towards processing the answers given to clarifying questions.
We hypothesize that not all information acquired through such interactions with the user would benefit the CS system, i.e., yield improvements in retrieval effectiveness.
Thus, the main novelty of our approach is that we do not blindly utilize the questions and the answers, but only when they are deemed to be useful.
Specifically, we focus on the task of conversational passage retrieval and design a classifier aimed at assessing usefulness of the asked clarifying question and the provided answer.
We utilize the question or the answer only if they are deemed useful, by appending them to the conversational history and employing a query rewriting method to attain a more information-dense query.
Results on the TREC 2022 Conversational Assistance Track (CAsT'22)~\cite{Owoicho:2022:TREC} demonstrate significant improvements in passage retrieval performance with the use of enhanced queries, as opposed to a non-mixed-initiative retrieval system ($12\%$ and $3\%$ relative improvement in terms of Recall@1000 and nDCG, respectively).
Further, when contrasting our approach to an established method that simply appends the prompted clarifying question and its answer to the original query~\cite{Aliannejadi:2021:EMNLP}, we observe differences in performance.
Specifically, if neither the question nor the answer are deemed useful, but still used, there is a relative performance decrease of $13\%$ in terms of nDCG@3, compared to non-mixed-initiative baselines.
In other words, it is better not to use any information provided by such questions and answers, than to use it wrongly.
Our contributions can be summarized as follows:
    \begin{itemize}
        \item We propose a simple, yet effective, method for processing answers to clarifying questions. The method is based on classifying usefulness of the prompted question and the given answer.
        \item We identify scenarios where asking clarifying questions resulted in improved passage retrieval, and where it decreased the retrieval performance.
    \end{itemize}

%% file: ecir2023-castmi-02.tex
\section{Related work}
\label{sec:related}

To facilitate further research in conversational search (CS), the TREC Conversational Assistance Track (CAsT)~\cite{Dalton:2020:TREC} aims to provide a reusable benchmark composed of several pre-defined conversational trajectories over a variety of topics.
The most recent edition, CAsT'22, offers a subtask oriented towards mixed-initiative (MI) interactions~\cite{Owoicho:2022:TREC}.
Under the MI paradigm of CS, the system can at any point of a conversation take initiative and prompt the user with various suggestions or questions~\cite{Radlinski:2017:CHIIR}.
One of the most prominent usages of mixed-initiative is asking clarifying questions with a goal of elucidating users underlying information need~\cite{Aliannejadi:2019:SIGIR}.
Recent research demonstrates a positive impact of clarifying questions both on user experience~\cite{Zamani:2020:SIGIR,Kiesel:2018:SIGIR} and retrieval performance~\cite{Aliannejadi:2021:EMNLP}.
Although other collections with clarifying questions and answers exist, most notably ClariQ~\cite{Aliannejadi:2021:EMNLP}, in this work, we focus on the aforementioned CAsT'22 as we additionally have control over which kind of question is being asked.
In general, two streams of approaches to constructing clarifying questions exist: (1) select an appropriate question from a pre-defined pool of questions~\cite{Aliannejadi:2019:SIGIR,Aliannejadi:2021:EMNLP,Owoicho:2022:TREC,Rosset:2020:WWW,Rao:2018:ACL}; (2) generate the question~\cite{Zamani:2020:WWW,Sekulic:2021:ICTIR,Majumder:2021:NAACL}.
However, despite the abundance of research on clarifying question construction, researched aimed at processing users' answers to such questions is scarce.
To bridge this gap, Krasakis et al.~\cite{Krasakis:2020:ICTIR} conduct an analysis of users' answers and find that they vary in polarity and length, as well as that retrieval effectiveness is often hurt.
Thus, in this work, we aim to automatically assess their usefulness, with a goal of mitigating this undesired effect.

%% file: ecir2023-castmi-03.tex
\section{Methodology}
\label{sec:method}

In this section, we formally define the task of conversational passage retrieval under the mixed-initiative (MI) paradigm and present our methods for each of the components of the task, i.e., query rewriting, clarifying question selection, answer processing, and passage retrieval.

\subsection{Task Formulation}
\label{sec:method:task}
At a current conversational turn $t$, given a user utterance $u^t$ and a conversation history $H=[(u^1, s^1),\dots,(u^{t-1}, s^{t-1})]$, the task is to generate a system response $s^t$.
For clarity, we omit the superscript $t$ from the subsequent definitions.
In MI CS systems, the system's response $s$ can either be a clarifying question $s_{cq}$ or a ranked list of passages $s_D$, $D = [d_1, d_2,\dots,d_N]$, where $N$ is the number of passages retrieved and $d_i$ is the $i$-th passage in the list.
Similarly, user utterance $u$ can take form of a query $u_q$ or an answer $u_a$ to system's question $s_{cq}$.
Modeling other types of user utterances, such as explicit feedback, falls out of the scope of this study.
Following prior work~\cite{Vakulenko:2021:ECIR}, the first task, i.e., query rewriting, is aimed towards resolution of the user query $u_q$ in the context of the conversation history, resulting in $u'_q = \gamma(u_q | H)$, where $\gamma$ is a query rewriting method.

Following the MI setting introduced at TREC CAsT'22~\cite{Owoicho:2022:TREC}, we address the problem of conversational passage retrieval through the following three components:
\begin{enumerate*}[label=(\roman*)]
    \item Produce system utterance $s_{cq}$ by selecting an appropriate clarifying question $cq$ from a given pool of questions $PQ$;
    \item Process the given answer $u_a$ and incorporate relevant information to the current query, resulting in $u''_q = \theta(u'_q, s_{cq}, u_a)$;
    \item Return a ranked list of passages $s_D$.
\end{enumerate*}
\noindent Next, we define our approaches to the described components.
We note that a clarifying question might be needed only for ambiguous, faceted, or unclear user requests.
Thus, for queries not requiring clarification, the system might opt to return a ranked list of passages without asking further questions.
However, following the setup enabled by CAsT'22 track, we do not explicitly model clarification need and thus design a system that prompts the user with a clarifying question at each turn.

\subsection{Clarifying Question Selection}
\label{sec:method:cq_filter}

For each query $u'_q$, we rank the potential candidates $cq_i$ based on their semantic similarity to $u'_q$.
Specifically, we utilize a T5 model fine-tuned on the CANARD dataset~\cite{Elgohary:2019:EMNLP}, available at HuggingFace,\footnote{\url{https://huggingface.co/castorini/t5-base-canard}} as our $\gamma$ rewriting function, which yields a resolved utterance $u'_q$.
To rank the potential candidates, we use MPNet~\cite{Song:2020:arXiv} from SentenceTransformers~\cite{Reimers:2019:EMNLP}, trained for semantic matching.
We select $cq_i$ with the highest score, as predicted by the MPNet: $s_{cq} = \mathrm{argmax}_{cq_i \in PQ_f} \mathit{MPNet}(u'_q, cq_i)$, where $PQ_f$ is a pool of clarifying questions with potentially misleading, unreliable, and faulty questions automatically filtered from the pool~\cite{Lajewska:2023:TREC}.

\subsection{Answer Processing}
\label{sec:method:answer}

\begin{table}[t]
\caption{Examples of annotated subset of ClariQ, indicating cases when clarifying question, answer, both, or neither are useful.}
\label{tbl:examples}
\begin{adjustbox}{max width=\textwidth}
\begin{tabular}{@{}llllr@{}}
\toprule
Query (initial request) & Clarifying Question & Answer & Useful & Prevalence \\
\midrule
\begin{tabular}[c]{@{}l@{}}I'm looking for information\\  on hobby stores.\end{tabular} & \begin{tabular}[c]{@{}l@{}}Do you want to know\\  hours of operation?\end{tabular} & No. & \textit{None} & 32\% \\
\begin{tabular}[c]{@{}l@{}}Tell me information about\\  computer programming.\end{tabular} & \begin{tabular}[c]{@{}l@{}}Are you interested in\\  a coding bootcamp?\end{tabular} & \begin{tabular}[c]{@{}l@{}}No, I want to know \\ what career options\\  programmers have\end{tabular} & \textit{Answer} & 53\% \\
Find me map of USA. & \begin{tabular}[c]{@{}l@{}}Do you want to see a map\\  of US territories?\end{tabular} & Yes. & \textit{Question} & 11\% \\
All men are created equal & \begin{tabular}[c]{@{}l@{}}Would you like to know more about\\ the declaration of independence?\end{tabular} & \begin{tabular}[c]{@{}l@{}}Yes, I'd like to\\ know who wrote it\end{tabular}  & \begin{tabular}[c]{@{}l@{}}\textit{Question} \\ and \textit{answer}\end{tabular} & 6\%
\\ \bottomrule
\end{tabular}
\end{adjustbox}
\end{table}

In this subsection, we describe our novel usefulness-based approach to processing answers given to the asked clarifying questions.
To address the issue, we move away from previous approaches that simply append the question and the answer to the original query~\cite{Aliannejadi:2019:SIGIR,Aliannejadi:2021:EMNLP}, regardless of the actual information gain.
In fact, a recent study by Krasakis et al.~\cite{Krasakis:2020:ICTIR} demonstrated that such practice can cause a decrease in retrieval effectiveness.
Moreover, they show that multi-word answers are informative (e.g., ``yes, I'm looking for info on spiders in Europe''), thus improving retrieval performance.
Similarly, short negative answers are not informative (e.g., ``no''), while multi-word negative answers are (e.g., ``no, I'm interested in buying aquarium cleaner'').
Thus, we define four possible actions, based on the current resolved utterance $u'_q$, the clarifying question asked $s_{cq}$, and the answer $u_a$:
\begin{enumerate}
    \item In case the answer is affirmative (e.g., ``yes'' or ``Yes, that is what I'm looking for''), we expand the $u_q'$ by appending the clarifying question asked.
    \item In case the answer is deemed useful, i.e., the underlying information need is explained in greater detail, we expand $u_q'$ by appending the answer.
    \item In case the answer is affirmative and it provides additional details, we expand $u_q'$ with both the clarifying question and the answer.
    \item If neither (1), (2), nor (3) is the case, we do not expand the utterance.
\end{enumerate}
Examples of the described cases are presented in Table \ref{tbl:examples} and are all aimed at incorporating additional useful information to the current utterance.
Formally:
\begin{equation}
    u''_q = 
    \left\{
    \begin{array}{cc}
       u'_q, & \quad \psi(u'_q, s_{cq}, u_a) = 0 \\
      \phi(u'_q, s_{cq}), & \quad  \psi(u'_q, s_{cq}, u_a) = 1 \\
      \phi(u'_q, u_a), & \quad \psi(u'_q, s_{cq}, u_a) = 2 \\
      \phi(u'_q, u_a, s_{cq}), & \quad \psi(u'_q, s_{cq}, u_a) = 3
   \end{array}
   \right.
\end{equation}
\noindent where $\psi(u'_q, s_{cq}, u_a)$ is a usefulness classifier, which aims to predict which of the above described actions to take.
The labels $0$, $1$, $2$, and $3$ correspond to neither $s_{cq}$ or $u_a$ were deemed useful, $s_{cq}$ was deemed useful, $u_a$ was deemed useful, and both were useful, respectively.
Similarly to Section~\ref{sec:method:task}, the function $\phi$ rewrites the original query given the context, in this case $s_{cq}$ or $u_a$.

Specifically, to model $\psi$, we fine-tune a large transformer-based model, namely T5~\cite{Raffel:2020:JMLR}, for multi-class classification.
To fine-tune the classifier, we manually annotate a portion of ClariQ ($150$ samples) for the specific aforementioned cases. The annotations were performed by two authors of the paper with an inter-annotator agreement Cohen's kappa of 0.89. The differences in annotations were discussed and resolved consensually.
Examples of annotations are presented in Table \ref{tbl:examples} and classification performance is reported in Section~\ref{sec:results:usefulness}.
We dub our novel mixed-initiative classifier-based method \emph{MI-Clf}.
Moreover, we assess the prevalence of each of the cases, and find, as presented in Table \ref{tbl:examples}, that 68\% of interactions contain new, useful information.
In the other 32\% of the cases, the answer simply negates the prompted clarifying question.
Although this interaction also provides valuable insights into the user's information need, current approaches would expand the query by appending the prompted clarifying question and the answer.
However, such an expanded query contains terms the user is not interested in, which can potentially degrade retrieval performance.
We compare our proposed method to such a baseline, which always extends the query as: $u''_q = \phi(u'_q, s{cq}, u_a)$.
This method is dubbed \emph{MI-All}.

\paragraph{Passage Retrieval.}
Finally, the rewritten utterance $u''_q$ is fed into a standard two-stage retrieve-and-rerank pipeline~\cite{Lajewska:2023:ECIR}.
We utilize BM25
($k1=0.95$, $b=0.45$) with RM3, where the initial query is extended with the highest-weighting terms from top-k scoring passages ($k=10$ and number of terms $m=10$).
Next, we use a point-wise monoT5 re-ranker~\cite{Nogueira:2020:EMNLP} to re-rank the top 1000, followed by a pair-wise duoT5 re-ranker~\cite{Pradeep:2021:arXiv} to additionally re-rank the top 50 passages.
The non-mixed-initiative baseline, dubbed \emph{DuoT5}, uses the same retrieval pipeline.

%% file: ecir2023-castmi-05.tex
\section{Results}
\label{sec:results}

\paragraph{Usefulness Classifier.}
\label{sec:results:usefulness}

The proposed usefulness classifier, described in Section~\ref{sec:method:answer}, achieves an average macro-$F_1$ score of $0.75$ and accuracy of $89\%$ in a stratified 5-fold evaluation on the aforementioned annotated subset of ClariQ. 
Next, we employ the trained classifier to predict the usefulness of ($u'_q$, $s_{cq}$, $u_a$) at each turn in the CAsT'22 dataset.
The question $s_{cq}$ was classified as useful in $28\%$ of turns, while the answer $u_a$ in $37\%$. 
In the rest $35\%$ of the cases, neither was predicted to be useful.
While the distribution of the predictions is similar to the prevalence in human-annotated data reported in Table~\ref{tbl:examples}, some differences can be observed.
For example, in CAsT'22 $28\%$ of the clarifying questions were deemed useful, as opposed to the $13\%$ in ClariQ.

\begin{table}[t]
	\caption{Performance of baselines and our mixed-initiative approaches on CAsT'22.} \label{tbl:main}
        \centering
	\begin{tabular}{lllllll}
        \toprule
		\textbf{Approach/RunID} & \textbf{R@1000} & \textbf{MAP} &
        \textbf{MRR} & \textbf{NDCG} & \textbf{NDCG@3} & \textbf{NDCG@5}
		\\ \midrule
		BM25\_T5\_automatic & 0.3244 & 0.1498 & 0.5272 & 0.2987 & 0.3619 & 0.3443 \\
		BM25\_T5\_manual & 0.4651 & 0.2309 & 0.7155 & 0.4228 & 0.5031 & 0.4831 \\
        \midrule
        our\_baseline (DuoT5) & 0.3846 & 0.1680 & 0.4990 & 0.3392 & 0.3593 & 0.3502 \\
        +MI-All & \textbf{0.4441} & 0.1741 & \textbf{0.5297} & 0.3594 & \textbf{0.3722} & 0.3508 \\
        MI-Clf & 0.4302 & \textbf{0.1776} & 0.5144 & \textbf{0.3613} & 0.3697 & \textbf{0.3581} \\
        \bottomrule
	\end{tabular}
\end{table}

\paragraph{Retrieval Performance.}
\label{sec:results:retrieval}
Results of the end-to-end conversational passage retrieval task, after the applied mixed-initiative answer processing methods (\emph{MI-All} and \emph{MI-Clf}) are presented in Table \ref{tbl:main}.
For reference, we also include the organizers' baselines in the table.
We make several observations from the presented results.
First, both methods that utilize mixed-initiative show improvements over the DuoT5 method. This confirms previous findings on the positive impact of clarifications in conversational search.
Second, differences between \emph{MI-All} and \emph{MI-Clf} are not statistically significant, across all metrics. However, we note that our classifier-based method utilizes clarifying question or the answer only when deemed useful, which is in about 70\% of the cases in CAsT'22. On the contrary, \emph{MI-All} always utilizes both the clarifying question and the answer. 
The equal performance of the two methods suggests that our usefulness classifier successfully includes only relevant information.

%% file: ecir2023-castmi-06.tex
\paragraph{Analysis.}
\label{sec:analysis}

In cases where the usefulness classifier predicted that neither the clarifying question $s_{cq}$ nor the answer $u_a$ is useful, we observe a drop of the \emph{MI-All} method's retrieval performance, in terms of nDCG@3 ($-13\%$).
Recall, however, is not impacted by incorporating potentially not useful information and even shows a slight increase ($+3.3\%$).
As this method always appends both $s_{cq}$ and $u_a$ to the query $u'_q$, the performance drop is expected, especially in the re-ranking stage, as the re-ranker might be confused by the additional non-relevant information.
Moreover, for both MI methods, we observe higher performance gains when the answer is useful ($+19.8\%$ for \emph{MI-All} and $+23.4\%$  for \emph{MI-Clf} in recall), compared to cases when the question is useful ($+12.5\%$ for \emph{MI-All} and $+8.5\%$ \emph{MI-Clf} in recall).
This could be explained by the fact that users' answers are deemed useful when they are longer and thus provide more detail on the underlying information need~\cite{Krasakis:2020:ICTIR}.
On the contrary, a clarifying question can be deemed useful even when tangibly addressing the user's need.
In other words, a good clarifying question can make a small step towards elucidating the user's information need.
However, the user's answer can contain detailed expression of their information need, thus making further gains.

%% file: ecir2023-castmi-07.tex
\section{Conclusion}
\label{sec:concl}
In this study, we proposed a classifier-based method, \emph{MI-Clf}, for processing answers to clarifying questions in conversational search, which extends the original query only when either is deemed useful.
Results on the TREC CAsT'22 dataset demonstrate clear improvements of the \emph{MI-Clf} method over non-mixed-initiative baselines ($+12\%$ and $+3\%$ relative improvement in terms of Recall@1000 and nDCG).
Moreover, we observed a performance drop for established methods that always use both the clarifying question and the answer, in cases where neither is useful ($-13\%$ in terms of nDCG@3), thus incorporating noisy information.
This study makes the first steps towards improved answer processing methods. 

\subsubsection{Acknowledgments}

This research was partially supported by the Norwegian Research Center for AI Innovation, NorwAI (Research Council of Norway, project number 309834).